\documentclass[traditabstract]{aa}

\usepackage{graphicx}
\usepackage{txfonts}
\usepackage{natbib}
\usepackage{longtable,pdflscape}
\usepackage{booktabs}
\usepackage{multirow}
\usepackage[breaklinks, colorlinks, citecolor=blue]{hyperref}
\usepackage[para,online,flushleft]{threeparttable}
\def\arcs{\ifmmode {^{\scriptstyle\prime\prime}}
          \else $^{\scriptstyle\prime\prime}$\fi}
\def\parcm{\sa=.08em \sb=.03em
     \ifmmode \hbox{\rlap{.}\kern\sa}^{\scriptstyle\prime}\hbox{\kern-\sb}
     \else \rlap{.}\kern\sa$^{\scriptstyle\prime}$\kern-\sb\fi}
\def\arcm{\ifmmode {^{\scriptstyle\prime}}
          \else $^{\scriptstyle\prime}$\fi}
\def\parcs{\sa=.07em \sb=.03em
     \ifmmode \hbox{\rlap{.}}^{\scriptstyle\prime\kern -\sb\prime}\hbox{\kern -\sa}
     \else \rlap{.}$^{\scriptstyle\prime\kern -\sb\prime}$\kern -\sa\fi}

\def\apjl{ApJL}

\begin{document}

\title{The dynamical state of RXCJ1230.7+3439: a multi-substructured merging galaxy cluster}
\titlerunning{The dynamical status of RXCJ1230 cluster of galaxies}

\author{R.~Barrena \inst{1,2} \and H. B\"ohringer \inst{3,4} \and G. Chon
\inst{4}}
\institute{Instituto de Astrof\'{\i}sica de Canarias, C/V\'{\i}a L\'{a}ctea s/n, E-38205 La Laguna, Tenerife, Spain\\
\email{rbarrena@iac.es} 
\and
Universidad de La Laguna, Departamento de Astrof\'{i}sica, E-38206 La Laguna, Tenerife, Spain
\and
Max-Planck-Institut f\"ur extraterrestrische Physik, D-85748 Garching, Germany
\and
Universit\"ats-Sternwarte M\"unchen, Fakult\"at f\"ur Physik,
Ludwig-Maximilian-Universit\"at M\"unchen, Scheinerstr. 1, D-81679 M\"unchen,
Germany}

\date{Received ; accepted } 

\authorrunning{Barrena et al.}

\abstract{We analyse the kinematical and dynamical state of the galaxy cluster RXCJ1230.7+3439 (RXCJ1230), 
at z$=0.332$, using 93 new spectroscopic redshifts of galaxies acquired at the Telescopio Nazionale
Galileo and from SDSS DR16 public data. We study the density galaxy distribution retrieved from photometric 
SDSS multiband data. We find that RXCJ1230 appears as a clearly isolated peak in the redshift space, 
with a global line-of-sight (LOS) velocity dispersion $\sigma_\textrm{v}=1004_{-122}^{+147}$ km s$^{-1}$. 
Several tests applied to the spatial and velocity distributions reveal that RXCJ1230 is a complex 
system with the presence of three subclusters, located at the south-west, east and south with respect to 
the main body of the cluster, containing several bright galaxies (BGs) in their respective cores. Our 
analyses confirm that the three substructures are in a pre-merger phase, where the main 
interaction takes place with the south-west subclump, almost in the plane of the sky. We compute a 
velocity dispersion of $\sigma_\textrm{v} \sim 1000$ and $\sigma_\textrm{v} \sim 800$ km s$^{-1}$ for the main 
cluster and the south-west substructure, respectively. The central main body and south-west 
substructure differ by $\sim 870$ km s$^{-1}$ in the LOS velocity. From these data, we estimate a 
dynamical mass of $M_{200}= 9.0 \pm 1.5 \times 10^{14}$ M$_{\odot}$ and $4.4 \pm 3.3 \times 
10^{14}$ M$_{\odot}$ for the RXCJ1230 main body and south-west clump, respectively, which reveals 
that the cluster will suffer a merging characterized by a 2:1 mass ratio impact. We solve a two-body 
problem for this interaction and find that the most likely solution suggests that the merging axis 
lies $\sim17^\circ$ from the plane of the sky and the subcluster will fully interact in $\sim0.3$ Gyr. 
However, a slight excess in the X-ray temperature observed in the south-west clump confirms a 
certain degree of interaction already. The comparison between the dynamical masses and those derived 
from X-ray data reveals a good agreement within errors (differences $\sim 15$\%), which suggests that the 
innermost regions ($<r_{500}$) of the galaxy clumps are almost in hydrostatical equilibrium. To summarize, 
RXCJ1230 is a young but also massive cluster in a pre-merging phase accreeting other galaxy systems 
from its environment.}


\maketitle

\section{Introduction}
\label{sec:intro}

According to the hierarchical structure formation scenario, Galaxy Clusters (GCs) are the youngest 
bound systems in our Universe. The cold dark matter model, together with the theory of cosmic 
inflation predicts the initial conditions for the structure formation \citep[see e.g.][]{Spri05}, 
where clusters form in the deepest potential wells generated by the dark matter (DM) 
overdensities. GCs contain multple components. In addition to DM, haloes include baryonic matter 
in different forms \citep[see e.g.][]{Alle11}. Cold and hot gas and non-thermal plasma 
constitute the intra-cluster medium (ICM), which highly reacts in collision processes, while the 
galactic component is less affected in mergings events. This multi-component nature allows us to 
analyse GCs through different wavelengths, given that different physical processes befell in each 
case. For example, we use X-rays and radio observations to probe the ICM \citep{Loew03,Boh10, 
vanWee19}, while visible and infrared data are used to analyse the galactic behaviour \citep[see][for 
a review]{Biv00}. On the other hand, DM is better studied through weak-lensing techniques 
\citep{Ume00}. Therefore, the multi-wavelength observations provide a more complete view about 
the reality of GCs.

X-ray observations are often used to investigate the dynamical state of GCs. % Indeed, X-ray 
% luminosity is tightly related to the cluster mass (ref??? e.g. Kravtsov et al. (2006; ApJ, 650, 
% 128); Pratt et al. (2009; A\&A, 498, 361)). 
However, optical information is an essential technique to study the dynamics of cluster mergers 
\citep{Golo19}. The spatial distribution (obtained from photometric observations) and kinematics 
(retrieved using spectroscopic redshifts) of galaxy members allow us to identify substructures 
and to analyse possible pre- and post-merging scenarios. Moreover, optical data complement the 
X-ray information because ICM and galaxies react on different time-scales during a collision 
\citep{Roe97}, thus the importance of combining X-ray and optical data. In this context, we 
analyse in this work the dynamical state of the RXCJ1230.7+3439 (hereafter RXCJ1230) cluster of 
galaxies using X-ray and optical data.

RXCJ1230 was discovered using X-ray data by \citet{Appe98} under the designation ATZ98-D219 and 
by \citet{Boh00} as part of the NORAS survey, with a redshift of z$=0.333$. This cluster has been 
also detected through its Sunyaev-Zeldovich (SZ) signal in the second Planck cluster catalog 
\citep{PC16}. It is also detected in the Northern Sky Cluster Survey \citep{Gal03} and in the 
Sloan Digital Sky Survey (SDSS) by \citet{Wen09} and \citet{Hao10}. RXCJ1230 contains a strong 
X-ray point source in the North of the cluster, associated with the NVSS 123050+344257 radio 
galaxy, at the redshift of the cluster.

To date, there is a poor spectroscopic information available in the literature and databases. 
For instance, only 15 spectroscopic redshifts are reported by SDSS-DR16 within a region of 
$15'$ radius with respect to the center of the cluster, insufficient to perform any 
detailed dynamical study. So, we have recently carried out multiobject spectroscopic 
observations at the TNG 3.5m telescope obtaining new redshift data for 81 galaxies in the 
field of RXCJ1230. In addition, we also include photometric information retrieved from SDSS-DR16 
and Pan-STARRS1 imaging archives. All these data, together with X-ray information will allow us 
to investigate the kinematics and dynamics of RXCJ1230, obtaining a satisfactory answer for
questions such as, is this a merging cluster?, is it in a pre- or a post- merger phase?, how well 
X-ray and dynamic mass estimates match?, ... Here, we will clarify the dynamical stage of RXCJ1230.

This paper is organized as follows. We describe the new spectroscopic observations and data samples 
in Sect. \ref{sec:optical_obs}. In Sect. \ref{sec:optical_ana} we present our results about the 
selection of galaxy members, velocity and spatial distributions, and substructures. In Sect. 
\ref{sec:dynamics}, we present global properties, dynamical masses of the different structures and
a comparison with X-ray properties. In Sect. \ref{sec:merging}, we analyse the 3D dynamics of the 
complex and we propose a plausible pre-merging scenario for RXCJ1230. At the end of this paper, 
in Sect. \ref{sec:conclusions} we summarize our results and expose our conclusions.

In this paper, we have worked assuming a flat cosmology with $\Omega_m=0.3$, $\Omega_\Lambda=0.7$ 
and H$_0=70$ $h_{70}$ km s$^{-1}$ Kpc$^{-1}$. Under this cosmology, 1 arcmin corresponds to 287 
$h_{70}^{-1}$ Mpc at the redshift of the cluster.

\section{Data sample}
\label{sec:optical_obs}

\subsection{Optical spectroscopy}
\label{sec:spectroscopy}

We performed multi-object spectroscopic (MOS) observations of RXCJ1230 at the 3.5m TNG telescope
in 2020 March, mapping a region of about $10^\prime\times10^\prime$ with two MOS masks 
and including a total of 92 slits. We used the DOLORES spectrograph and the LR-B 
grism\footnote{See http://www.tng.iac.es/instruments/lrs},
which offers a wavelength coverage between 370 and 800 nm with a spectral resolution of 
2.75 \AA $\ $ per pixel. The total integration time was 3600 s per mask, divided in two
exposures of 1800 s each. The combination of these two accquisitions allowed us to
correct the spectra from cosmic rays.

The spectra were extracted using standard {\tt IRAF}\footnote{{\tt IRAF} 
(\url{http://iraf.noao.edu/}) is distributed by the National Optical Astronomy Observatories, 
which are operated by the Association of Universities for Research in Astronomy, Inc., under 
cooperative agreement with the National Science Foundation.} packages. The radial velocity 
computation was performed using the cross-correlation technique developed by \citet{Ton79} 
and implemented as the task {\tt RVSAO.XCSAO} in {\tt IRAF} environment. This procedure 
correlates the features detected in the observed spectra (mainly Ca H and K doublet, 
H$_\delta$, G band, MgI, in absorption and the most relevant emission lines such as OII, 
OIII doublet, H$_\alpha$, H$_\beta$  and H$_\delta$) with that present in the templates 
spectra. We used five different reference spectra of the Kennicutt Spectrophotometric Atlas 
of Galaxies \citep{ken92}, for 5 different morphologies (Elliptical, Sa, Sb, Sc and Irr 
types). So, this procedure yields radial velocity estimates and the corresponding errors 
due to the correlation technique applied. After rejecting the spectra with lower 
signal-to-noise ratio (SNR), we obtained 81 spectroscopic redshifts. In addition, we also 
consider 12 redshifts retrieved from the SDSS DR16 spectroscopic database present in the 
region sampled by the two masks field of view around RXCJ1230.

A detailed comparison between redshifts derived from multiple measures of the same target
(obtained from two different estimation in the two masks, or even between SDSS redshift and 
our estimate), we see that the nominal velocity errors provided by the cross-correlation 
technique are too small. Therefore, in order to convert this error into a realistic value, 
considering systematic errors, we need to multiply them by a factor of 2 \citep[see e.g.][]{Bos13}.

% \begin{landscape}
\begin{table}
\fontsize{9}{11}
% \begin{threeparttable}
\caption{Velocity catalogue of 93 galaxies measured spectroscopically in the
RXJ1230 field.}
\label{tab:catalog}
\tiny
\begin{tabular}{l c c p{5mm} p{5mm} l}
\midrule \midrule
 ID & R.A. \& Dec. (J2000)              & v$\pm \Delta$v & \makebox[8mm][c]{$r^\prime$} & \makebox[8mm][c]{$i^\prime$} & Notes \cr
    & R.A.=$12\! : \! mm \! : \! ss.ss$ & (km s$^{-1}$)  &            &            & \cr
    & Dec.=$+34\! :\! mm \! : \! ss.s$  &                &            &            & \cr
\midrule
1$^\star$  & 30:17.02  \enspace 34:22.3 & 98389  $\pm$ 102 & 19.69 & 19.25 & \cr
2          & 30:19.11  \enspace 40:59.0 & 90506  $\pm$ 77  & 19.78 & 19.22 & \cr
3$^\star$  & 30:22.18  \enspace 36:37.2 & 99651  $\pm$ 62  & 18.86 & 18.22 & \cr
4$^\star$  & 30:22.86  \enspace 36:44.4 & 100135 $\pm$ 79  & 20.66 & 20.14 & \cr
5$^\star$  & 30:23.72  \enspace 39:40.5 & 97351  $\pm$ 94  & 19.21 & 18.77 & ELG \cr
6          & 30:23.17  \enspace 40:41.3 & 118366 $\pm$ 76  & 19.55 & 19.12 & ELG \cr
7$^\star$  & 30:24.45  \enspace 40:01.7 & 97405  $\pm$ 82  & 20.63 & 20.07 & \cr
8          & 30:25.53  \enspace 40:28.5 & 90865  $\pm$ 85  & 19.04 & 18.43 & ELG \cr
9$^\star$  & 30:26.53  \enspace 36:01.1 & 100299 $\pm$ 68  & 20.39 & 19.89 & \cr
10$^\star$ & 30:26.00  \enspace 37:23.0 & 100012 $\pm$ 103 & 21.04 & 20.47 & \cr
11$^\star$ & 30:27.68  \enspace 40:49.8 & 98300  $\pm$ 100 & 20.75 & 20.33 & \cr
12$^\star$ & 30:27.16  \enspace 37:30.3 & 98085  $\pm$ 31  & 20.22 & 19.68 & \cr
13$^\star$ & 30:28.11  \enspace 34:51.2 & 98441  $\pm$ 142 & 20.57 & 20.16 & \cr
14$^\star$ & 30:28.03  \enspace 35:42.0 & 99284  $\pm$ 78  & 21.07 & 20.75 & \cr
15$^\star$ & 30:28.63  \enspace 37:35.5 & 99494  $\pm$ 60  & 19.73 & 19.16 & \cr
16$^\star$ & 30:28.61  \enspace 37:41.4 & 100177 $\pm$ 85  & 19.53 & 18.97 & \cr
17$^\star$ & 30:28.36  \enspace 37:52.8 & 97806  $\pm$ 98  & 20.80 & 20.37 & \cr
18$^\star$ & 30:28.78  \enspace 40:12.7 & 98013  $\pm$ 76  & 20.46 & 19.87 & \cr
19         & 30:28.96  \enspace 41:20.4 & 118244 $\pm$ 55  & 20.15 & 19.52 & ELG \cr
20$^\star$ & 30:29.28  \enspace 38:01.3 & 98445  $\pm$ 70  & 20.46 & 19.83 & \cr
21$^\star$ & 30:29.54  \enspace 38:06.9 & 97959  $\pm$ 16  & 18.07 & 17.50 & BG-W (2) \cr
22$^\star$ & 30:30.55  \enspace 38:01.4 & 98007  $\pm$ 18  & 17.90 & 17.33 & BG-W (1) \cr
23         & 30:31.78  \enspace 41:28.7 & 90405  $\pm$ 145 & 20.24 & 19.61 & \cr
24$^\star$ & 30:32.80  \enspace 38:11.5 & 98538  $\pm$ 92  & 20.59 & 20.10 & \cr
25$^\star$ & 30:33.48  \enspace 38:19.5 & 100904 $\pm$ 116 & 19.95 & 19.51 & \cr
26$^\star$ & 30:33.73  \enspace 38:32.7 & 98418  $\pm$ 54  & 19.53 & 19.03 & \cr
27$^\star$ & 30:35.37  \enspace 38:27.4 & 98488  $\pm$ 86  & 20.60 & 20.17 & \cr
28         & 30:36.83  \enspace 38:41.1 & 139900 $\pm$ 100 & 21.98 & 21.71 & ELG \cr
29$^\star$ & 30:37.75  \enspace 36:16.5 & 101460 $\pm$ 85  & 19.92 & 19.37 & \cr
30$^\star$ & 30:38.15  \enspace 34:58.3 & 99530  $\pm$ 100 & 20.70 & 20.21 & \cr
31$^\star$ & 30:38.52  \enspace 35:48.8 & 100702 $\pm$ 118 & 20.72 & 20.14 & \cr
32$^\star$ & 30:38.80  \enspace 39:55.4 & 102345 $\pm$ 100 & 19.86 & 19.20 & \cr
33$^\star$ & 30:38.19  \enspace 36:18.8 & 98484  $\pm$ 34  & 20.67 & 20.09 & \cr
34$^\star$ & 30:38.21  \enspace 38:06.3 & 100291 $\pm$ 25  & 19.92 & 19.38 & \cr
35         & 30:39.41  \enspace 34:36.2 & 86666  $\pm$ 111 & 21.04 & 20.55 & \cr
36$^\star$ & 30:39.73  \enspace 39:22.6 & 99681  $\pm$ 51  & 20.28 & 19.68 & \cr
37         & 30:39.18  \enspace 35:55.6 & 156900 $\pm$ 100 & 21.89 & 21.63 & ELG \cr
38$^\star$ & 30:39.97  \enspace 39:32.3 & 101073 $\pm$ 115 & 19.72 & 19.14 & \cr
39$^\star$ & 30:40.75  \enspace 35:29.6 & 99991  $\pm$ 93  & 20.25 & 19.63 & \cr
40$^\star$ & 30:41.59  \enspace 34:12.3 & 98373  $\pm$ 81  & 19.33 & 18.94 & \cr
41$^\star$ & 30:41.64  \enspace 39:51.4 & 99151  $\pm$ 67  & 20.61 & 20.13 & \cr
42         & 30:41.18  \enspace 41:48.5 & 112903 $\pm$ 63  & 20.32 & 20.12 & ELG \cr
43$^\star$ & 30:42.20  \enspace 41:40.3 & 101134 $\pm$ 182 & 19.91 & 19.33 & \cr
44$^\star$ & 30:42.48  \enspace 36:13.5 & 98196  $\pm$ 89  & 20.39 & 19.85 & \cr
45         & 30:42.48  \enspace 36:30.5 & 157618 $\pm$ 100 & 20.92 & 20.19 & ELG \cr
46$^\star$ & 30:42.57  \enspace 37:31.2 & 99913  $\pm$ 157 & 20.53 & 20.01 & \cr
47$^\star$ & 30:42.12  \enspace 38:01.7 & 99099  $\pm$ 61  & 20.21 & 19.67 & \cr
48$^\star$ & 30:42.47  \enspace 40:24.3 & 100238 $\pm$ 98  & 19.70 & 19.22 & \cr
49         & 30:43.81  \enspace 35:12.0 & 103864 $\pm$ 68  & 18.19 & 17.72 & \cr
50$^\star$ & 30:43.73  \enspace 36:27.2 & 100334 $\pm$ 105 & 20.35 & 19.76 & \cr
51$^\star$ & 30:43.56  \enspace 36:06.1 & 100379 $\pm$ 107 & 20.55 & 20.12 & \cr
52$^\star$ & 30:43.93  \enspace 39:21.2 & 100007 $\pm$ 35  & 20.47 & 19.94 & \cr
53$^\star$ & 30:44.95  \enspace 38:46.7 & 99914  $\pm$ 107 & 20.58 & 20.07 & \cr
54         & 30:44.83  \enspace 37:50.8 & 112918 $\pm$ 101 & 20.07 & 19.80 & ELG \cr
55$^\star$ & 30:44.47  \enspace 39:38.7 & 100638 $\pm$ 87  & 19.49 & 18.87 & \cr
56$^\star$ & 30:44.82  \enspace 40:58.1 & 99794  $\pm$ 62  & 20.17 & 19.52 & \cr
57$^\star$ & 30:45.65  \enspace 40:18.0 & 99754  $\pm$ 88  & 20.80 & 20.28 & \cr
58$^\star$ & 30:45.57  \enspace 35:41.8 & 102128 $\pm$ 68  & 20.45 & 19.90 & \cr
59$^\star$ & 30:45.92  \enspace 38:19.0 & 100622 $\pm$ 70  & 19.60 & 19.06 & \cr
60$^\star$ & 30:45.78  \enspace 39:26.3 & 100064 $\pm$ 28  & 18.49 & 17.89 & BCG  \cr
61$^\star$ & 30:47.98  \enspace 36:56.7 & 99109  $\pm$ 78  & 18.69 & 18.10 & BG-S  \cr
62$^\star$ & 30:47.93  \enspace 37:03.7 & 96570  $\pm$ 100 & 20.80 & 20.74 & ELG \cr
63$^\star$ & 30:47.78  \enspace 43:26.5 & 100517 $\pm$ 86  & 19.54 & 18.98 & \cr
64$^\star$ & 30:48.41  \enspace 37:11.3 & 99457  $\pm$ 77  & 20.32 & 19.75 & \cr
65$^\star$ & 30:48.35  \enspace 36:48.7 & 99042  $\pm$ 108 & 21.82 & 21.26 & \cr
66$^\star$ & 30:48.52  \enspace 36:58.9 & 99102  $\pm$ 145 & 21.27 & 20.93 & \cr
67$^\star$ & 30:48.83  \enspace 37:07.7 & 99103  $\pm$ 55  & 20.68 & 20.17 & \cr
68$^\star$ & 30:48.50  \enspace 38:11.8 & 99959  $\pm$ 83  & 19.55 & 19.06 & \cr
69$^\star$ & 30:48.41  \enspace 38:47.2 & 102209 $\pm$ 60  & 21.19 & 20.52 & \cr
70$^\star$ & 30:48.66  \enspace 39:08.4 & 99497  $\pm$ 63  & 21.17 & 20.57 & \cr
71$^\star$ & 30:48.60  \enspace 42:27.7 & 99761  $\pm$ 63  & 19.60 & 19.08 & \cr
\midrule
\end{tabular}
% \end{threeparttable}
\end{table}
% \end{landscape}

% \begin{landscape}
\addtocounter{table}{-1}
\begin{table}
\fontsize{9}{11}
% \begin{threeparttable}
\caption{Continued.}
\tiny
\begin{tabular}{l c c p{5mm} p{5mm} l}
\midrule \midrule
 ID & R.A. \& Dec. (J2000)              & v$\pm \Delta$v & \makebox[8mm][c]{$r^\prime$} & \makebox[8mm][c]{$i^\prime$} & Notes \cr
    & R.A.=$12\! : \! mm \! : \! ss.ss$ & (km s$^{-1}$)  &            &            & \cr
    & Dec.=$+34\! :\! mm \! : \! ss.s$  &                &            &            & \cr
\midrule
72         & 30:48.06  \enspace 42:37.2 & 63400  $\pm$ 100 & 18.78 & 18.44 & \cr
73$^\star$ & 30:49.14  \enspace 37:17.7 & 98558  $\pm$ 88  & 20.36 & 19.98 & \cr
74$^\star$ & 30:49.72  \enspace 40:33.5 & 99387  $\pm$ 52  & 19.72 & 19.07 & \cr
75$^\star$ & 30:49.81  \enspace 42:10.1 & 101085 $\pm$ 77  & 21.03 & 20.39 & \cr
76$^\star$ & 30:49.43  \enspace 41:25.4 & 99249  $\pm$ 37  & 20.57 & 20.03 & \cr
77$^\star$ & 30:50.36  \enspace 42:01.2 & 101520 $\pm$ 100 & 20.58 & 20.11 & \cr
78$^\star$ & 30:50.67  \enspace 42:53.5 & 100907 $\pm$ 98  & 18.51 & 17.93 & NVSS RG \cr
79         & 30:51.83  \enspace 36:20.4 & 90272  $\pm$ 145 & 21.20 & 20.68 & \cr
80$^\star$ & 30:51.66  \enspace 38:59.0 & 101592 $\pm$ 49  & 19.63 & 18.94 & ELG \cr
81$^\star$ & 30:51.07  \enspace 40:48.1 & 100908 $\pm$ 71  & 20.73 & 20.18 & \cr
82$^\star$ & 30:51.03  \enspace 38:27.4 & 96834  $\pm$ 39  & 20.35 & 19.83 & \cr
83$^\star$ & 30:52.32  \enspace 43:18.4 & 100868 $\pm$ 97  & 20.12 & 19.64 & \cr
84$^\star$ & 30:53.33  \enspace 41:54.3 & 102042 $\pm$ 90  & 19.23 & 18.81 & ELG \cr
85$^\star$ & 30:54.56  \enspace 37:42.9 & 101422 $\pm$ 92  & 20.90 & 20.52 & \cr
86         & 30:54.99  \enspace 41:23.5 & 40650  $\pm$ 100 & 21.49 & 21.04 & \cr
87$^\star$ & 30:55.30  \enspace 41:31.8 & 99746  $\pm$ 48  & 18.68 & 18.10 & \cr
88         & 30:56.48  \enspace 41:46.3 & 92621  $\pm$ 165 & 20.46 & 19.77 & \cr
89$^\star$ & 30:57.96  \enspace 39:23.5 & 101676 $\pm$ 54  & 18.56 & 17.92 & \cr
90$^\star$ & 30:59.34  \enspace 39:13.9 & 100226 $\pm$ 19  & 19.51 & 18.92 & \cr
91$^\star$ & 31:01.05  \enspace 38:09.1 & 97834  $\pm$ 17  & 18.54 & 17.93 & \cr
92$^\star$ & 31:03.70  \enspace 39:08.1 & 99064  $\pm$ 20  & 18.64 & 18.05 & \cr
93$^\star$ & 31:04.67  \enspace 40:08.7 & 99113  $\pm$ 45  & 18.13 & 17.54 & BG-E  \cr
\midrule
\end{tabular}
% \end{threeparttable}
\footnotesize{Note: asterisk in column 1 (ID) indicates the galaxies selected as 
cluster members.}
\end{table}
% \end{landscape}

Our spectroscopic catalogue includes a total of 93 galaxy redshifts in a region of $9'\times10'$ 
arcmin (see Fig. \ref{fig:contours}, left panel). The MOS mask design was initialy planned to cover 
more efficiently the densest galaxy regions. So, the majority of the spectroscopic redshifts 
follows an elongated region in the NE-SW direction. Our full redshift sample presents a 
median SNR and $cz$ error of 8 and 82 km s$^{-1}$, respectively. We detect 12 star forming 
galaxies, characterized by the presence of the $[$OII$]$ emission line. 

\subsection{Optical photometry}
\label{sec:photometry}

We complement our redshift sample with photometric information retrieved from SDSS DR16
database. We consider $dered$ $r'$ and $i'$ magnitudes\footnote{$dered$ magnitudes are the 
extinction-corrected values following \citet{Sch98} reddening maps.}. Galaxy counts in this region
reveals the photometric SDSS sample is $\sim$90\% complete for galaxies down to magnitude 
$r'=21.5$, which is in agreement with the mean SDSS DR12 depth 
estimations\footnote{see https://www.sdss.org/dr12/imaging/other\_info/}.

Pan-STARRS1 data archive\footnote{see https://ps1images.stsci.edu/cgi-bin/ps1cutouts} 
was only used in order to retrieve RGB images and overplot density contours and galaxies 
with spectroscopic redshifts (see Fig. \ref{fig:contours}).

Comparing the spectroscopic and photometric samples in the area covered by MOS masks, we 
find that the completness of the spectroscopic sample is $\sim55$\% for galaxies down to 
magnitude $r'=21$. However, we are able to obtain redshifts even for galaxies with $r'>21.5$.

Table \ref{tab:catalog} lists the full spectroscopic sample considered in this work (see 
also Fig. \ref{fig:contours}). Col. 1 lists an ID number (cluster members are indicated), 
Cols. 2 and 3 reports the equatorial coordinates of galaxies in J2000 system, Col. 4 the 
heliocentric radial velocity ($v=cz$) with errors ($\Delta v$), and Cols 5 and 6, the 
completemtary photometric information ($r'$ and $i'$ $dered$ magnitudes) retrieved from 
SDSS DR16. The last column includes some comments regarding particular features of some galaxies.

\subsection{Complementary X-ray data}
\label{sec:x-ray}

The XMM-Newton observations (ID 0841900101) and their data analysis, which complete our 
multiwavelength study, are described in detail in \citet{Boh22}. Here, we only use X-ray
data for a qualitative and morphological analysis. We use the 0.5 to 2.0 keV band, which 
provides the highest signal to noise above the background. In the present work we perform 
two minor modifications to the raw X-ray image. First, we remove the point sources and 
emission from NVSS 123050+344257 radio galaxy (ID 78) by masking them with circular apertures 
of 10-20 pixels radius; Second, we smooth the original image using a Gaussian filter with 
a FWHM=12 arcsec. After this process, we obtain the X-ray surface brightness maps plotted as
contours in the right panel of Fig. \ref{fig:contours}, overlaid to an optical colour 
composite image from the PanSTARSS survey (similarly as shown in Fig. 2 by 
\citeauthor{Boh22} \citeyear{Boh22}). 

Analysis of the X-ray spectra in the different substructures of the cluster yields an 
intracluster plasma temperature of $4.7 \pm 0.4$ keV for the central main component, 
$4.4 \pm 0.6$ keV for the south-western subcluster and $3.3 {+0.7 \atop -0.6}$ keV for the 
eastern structure \citep{Boh22}. They also find that the temperature profile for the main 
component is not very steep with a polytropic index between isothermal and 1.2 and all 
components have no cool cores. From the temperature and the shape of the intracluster 
gas distribution \citet{Boh22} obtained a hydrostatic mass estimate of the cluster 
components. They combined this result with mass estimates based on scaling relations of 
cluster mass with X-ray temperature, with X-ray luminosity, with total gas mass and with 
$Y_X$ (the product of temperature and gas mass) to arrive at a consistent picture for 
the mass estimation. See Sect. \ref{sec:optical_xray_mass} for a summary of X-ray mass 
and a comparison with the dynamical one.

\section{Analysis and results}
\label{sec:optical_ana}

\subsection{Member selection and global properties}
\label{sec:members_global}

To select cluster members out of the 93 galaxies in our spectroscopic sample, we follow a method
based on the galaxy position in the 2D projected phase space ($r$, $cz$), where $r$ is the 
projected distance from the cluster center, and $cz$ is the galaxy line-of-sight velocity 
(see Fig. \ref{fig:3plots}, top panel). 

To minimize the presence of interlopers, we apply a 2.7$\sigma_\textrm{v}$ clipping in the $cz$ 
coordinate, taking into account the radial profile of the expected velocity dispersion \citep{Mam10}. 
Therefore, we apply an iterative method which, in a first step, we find the mean significant 
peak in the velocity distribution ($\textrm{v}_0=<cz_0>$) and estimate first velocity dispersion 
($\sigma_0=\sigma_{v,0}$) using the $rms$ estimator. Considering these two values, in a second 
step, we select cluster members as galaxies with $\textrm{v}< \textrm{v}_0\pm2.7\sigma_0$. In 
the final step, we refine the estimation of the mean $cz$ of the cluster and re-evaluate the 
velocity dispersion $\sigma_\textrm{v}$ of the cluster. This simple three-step procedure yields 
stable and converging values of $\bar{\textrm{v}}$ and $\sigma_\textrm{v}$ in a fourth and 
subsequent steps. Fig. \ref{fig:histogram} shows the redshift distribution of the galaxies 
listed in Tab. \ref{tab:catalog}.

\begin{figure*}[!h]
\centering
\includegraphics[]{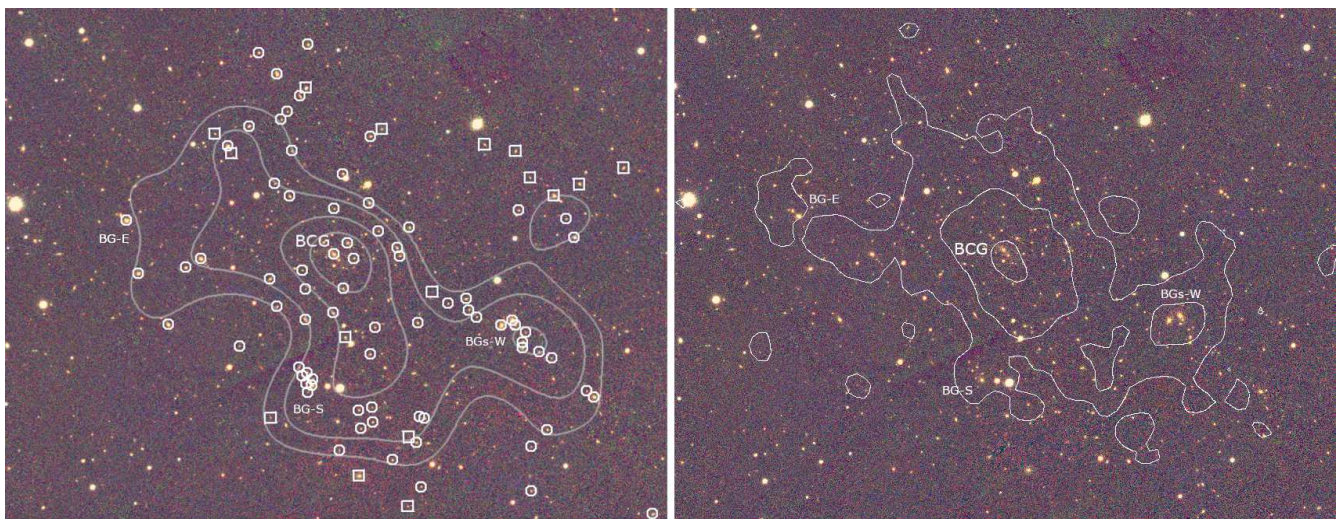}
\caption{{\it Left panel:} RGB colour composite image obtained by combining $g'-$, 
$r'-$ and $i'-$band images of $13^\prime\times10^\prime$ field of view from Pan-Starrs1 
public archive. Circles and squares correspond to galaxy members and non-members, 
respectively, obtained from our spectroscopic observations and SDSS-DR16 spectroscopic 
database. Superimposed, we also show the contour levels of isodensity galaxy distribution 
of likely members (see Sect. \ref{sec:spatial_distrib}). {\it Right panel:} The same RGB 
image but overplotting the contour levels of the XMM-Newton image corresponding 
to the observation ID 0841900101. The X-ray contours have been obtained after smoothing 
the original image using a Gaussian filter with $\sigma=6$ arcsec. Point sources and 
emission from NVSS 123050+344257 radio galaxy (ID 78) have been removed masking them 
with circular apertures of 10-20 pixels radius. In both panels, the BCG and bright galaxies 
(BGs) are also marked. North is up and East is left.}
\label{fig:contours}
\end{figure*}

\begin{figure}[h!]
\centering
\includegraphics[width=9cm,height=5cm]{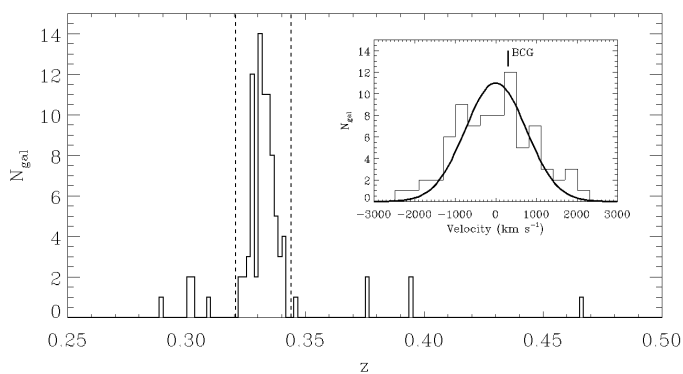}
\caption{Galaxy redshift distribution. Dashed vertical lines delimit the redshift
range including 77 galaxy members assigned to RXCJ1230 according to 2.7$\sigma_\textrm{v}$
clipping. Superimposed, the velocity distribution, in the cluster rest frame, of the 77 
cluster members selected. Black curve represents the reconstruction of the velocity 
distribution as a Gaussian profile, considering the $\sigma_v$ computed using the biweight 
estimator and assuming all the galaxies belonging to a single system. The velocity 
corresponding to the BCG is also marked.}
\label{fig:histogram}
\end{figure}

In this way, we select 77 galaxy members with a $\bar{\textrm{v}}=99658 \pm 161$ km s$^{-1}$ ($z=0.3324$) 
with an $rms$ of $969\pm 130$ km s$^{-1}$ (errors at 95\% c.l.) in the cluster rest frame. In order 
to verify these values, we also apply the bi-weight scale estimator \citep{Bee90} considering the 
77 redshifts. So, we find a $\sigma_\textrm{v}=1004_{-122}^{+147}$ km s$^{-1}$. Thus, both $rms$ and 
bi-weight estimators produce results in perfect agreement within errors. However, in order to 
check how robust this estimate is, we study the variation of $\sigma_\textrm{v}$ with the distance to the 
center of the cluster, which is taken to be the brightest cluster galaxy (BCG\footnote{Note that the 
BCG is not strictly the brightest galaxy of this complex cluster. The BCG is the brightest galaxy in the main body 
and more massive clump of galaxies, and lies very close to the X-ray peak emission and the maximum of 
galaxy isodensity distribution (see Fig. \ref{fig:contours}). There are other four galaxy members even 
brighter than the BCG in RXCJ1230 (see Tab. \ref{tab:catalog} and Figs. \ref{fig:contours} and 
\ref{fig:cmr}), lying in the surrounding subclusters. These four bright galaxies are labeled as BGs in order 
to differentiate them from the actual BCG. Probably, these four galaxies can also be considered as BCGs 
of their respective subclusters, but we label them as BGs to keep the notation more clear and a less 
confusing reading.}) position. Fig. \ref{fig:3plots}, bottom panel, shows that the integral 
$\sigma_\textrm{v}$ profile is flat beyond 0.7 Mpc, suggesting that the estimation of the $\sigma_\textrm{v}$ 
is robust for the whole cluster. In the following analyses we use the bi-weight estimator given its 
robustness in cases where the statistics clearly departs from the Gaussian distribution. In addition to 
the 77 members, we detect 6 galaxies in the foreground and 10 in the background of RXCJ1230.

Another remarkable effect that we find is a clear dependence of the mean velocity with the 
clustercentric distance (see Fig. \ref{fig:3plots}, middle panel). We notice that the inner and 
central region shows higher mean velocity, while regions surrounding the main body exhibit velcities
up to 500 km s$^{-1}$ lower.

The BCG of RXCJ1230 (the ID 60) presents a velocity of 100064$\pm$28 km s$^{-1}$, which is about 310 
km s$^{-1}$ higher respect to the mean velocity of the cluster. In addition to the BCG, with magnitude 
$r'=18.49$, we also detect two galaxies even brighter (IDs 22 and 21, with $r'=17.90$ and 18.07) located 
at the south-west. Besides these bright galaxies (BGs), we also identify two galaxies more, 
very bright and showing BCG features. One of them is located to the south (the ID 61, with $r'=18.69$), and 
another to the east (the ID 93, with $r'=18.13$). As Fig. \ref{fig:contours} shows, the BCG and all BGs are 
very close to the X-ray peaks as recovered from XMM-Newton data. The BCG is coincident with the main (and 
central) body of the cluster, while the BG-W(1) and BG-W(2) are very close to the south-west X-ray peak, 
In a similar way, the BG-S is almost coincident with a small elongation of the X-ray surface brightness 
toward the south, and the BG-E is placed in the maximum of a weak X-ray emission located to the east of the 
cluster. As we discuss in following sections, each one of these BGs is linked to a corresponding 
substructure, configuring a complex multi-substructure cluster. 

Additionally, to this set of BGs, we detect 12 galaxies showing [OII] emission lines,
but only 4 of these galaxies are cluster members. The SNR and the spectral resolution of
our data allow us to detect [OII] lines with equivalent width >8 \AA. These 
emission line galaxies (ELGs; star-forming galaxies) are the IDs 5, 62, 80 and 84, and 
show [OII] equivalent width of 15, 95, 12 and 18 \AA, respectively. The ELG  
members represent the 5.2\% of the cluster members in our sample. This fraction indicates 
that the star-forming processes have been quenched in RXCJ1230, which is in agreement 
with a presence of high galaxy density environments with ICM showing high $T_X$ 
\citep{Lag08}.

\subsection{Velocity field}
\label{sec:velocity}

Deviations from Gaussianity in the radial velocity distributions are clear indicators that
clusters present substructures \citep[see e.g.][]{Rib11}. In order 
to check whether the velocity distribution of RXCJ1230 follows a Gaussian shape, we use two
profile estimators, the skewness and kurtosis indexes. Positive skewness indicates the 
distribution is skewed to the right, with a longer tail to the right of the distribution 
maximum, while negative skewness indicates that the distribution is shifted and tailed to the 
left. On the other hand, positive values of the kurtosis indicate distributions presenting 
thinner tails (leptokurtic) than the normal distribution, while negative values indicate 
distributions with fatter tails (platykurtic). In our case, we obtain -0.038 and -0.532 
for the skewness and kurtosis, respectively. These values suggest that the velocity distribution 
follows a slighly flatter shape than a normal one, and it is quite symmetric. This implies 
that probably most of the substructures are placed close to the plane of the sky and no 
significant velocity deviations are expected in the radial component (along the line of sight).

\begin{figure}[h!]
\centering
\includegraphics[width=9cm,height=14.3cm]{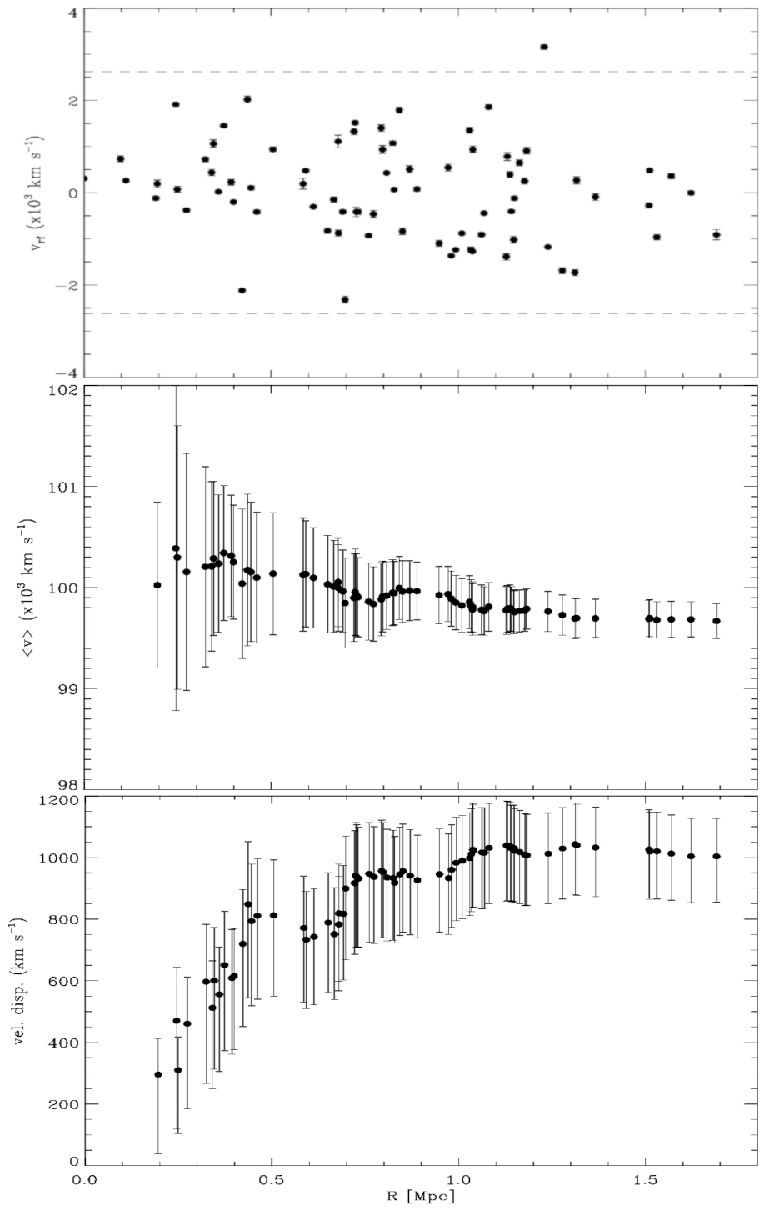}
\caption{Top panel: rest frame velocity versus projected distance to the cluster
centre for the 77 galaxy members selected. The cluster centre is assumed to be the
position of the BCG. Middle and bottom panels: Integral profiles and LOS
velocity dispersion, respectively. These values are computed by considering all
galaxies within that radius. The first value computed is estimated from the first
five galaxies closest to the centre. The error bars are at the 68\% c.l.}
\label{fig:3plots}
\end{figure}

\subsection{2D galaxy distribution}
\label{sec:spatial_distrib}

\begin{figure}[h!]
\centering
\includegraphics[width=9cm,height=6cm]{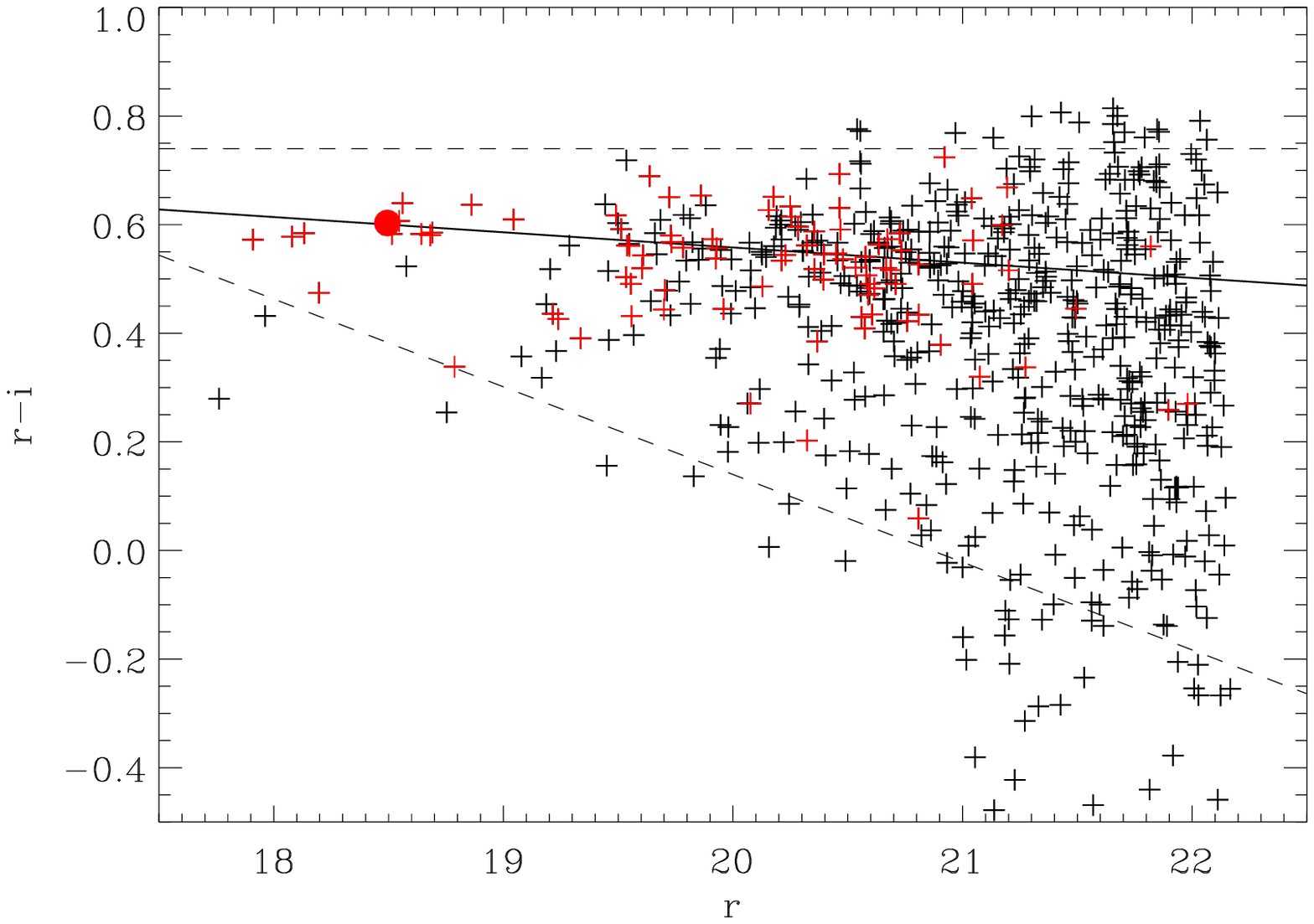}
\caption{Colour magnitude diagram ($r'-i'$,$r'$) of galaxies in a region of 
$12.4' \times 9.6'$. Red symbols correspond to galaxy members confirmed 
spectroscopically (red dot corresponds to the BCG of the cluster). Solid line 
represents the red sequence defined as the densest locus in this diagram, which
follows the linear fit $r'-i'=-0.028*r'+1.118$. Dashed lines delimit the region 
which enclose the RS and the blue cloud in this diagram. Galaxies included in 
this region are considered likely members, which are used to obtain the 
isodensity galaxy distribution shown in Fig. \ref{fig:contours}.}
\label{fig:cmr}
\end{figure}

Given that our spectroscopic sample suffers from magnitude incompleteness and does not map the 
whole cluster field, we also adopt the photometric SDSS DR16 catalogues. Using the $r'$ and 
$i'$ SDSS DR16 photometry in a region of $12.4' \times 9.6'$, we construct the ($r'-i'$ vs $r'$) 
colour magnitude relation (CMD; see Fig. \ref{fig:cmr}) and select likely members from there. 
First, we fit a red sequence (RS) to the spectroscopicaly confirmed members by fixing the 
slope to -0.028 following prescription detailed in \citet{Bar12}. So, we obtain $r'-i'=-0.028 
\times r'+1.118(\pm0.05)$. Then, in order to select both likely early-type members (placed in 
the RS) and galaxy members residing in the green valley and blue-cloud (below the RS) in the CMD 
of the cluster \citep{Eal18}, we select the locus defined by the RS$\pm 3 \times rms$ as upper 
limit and $-0.1615 \times r'+3.37$ as lower limit in $r'-i'$, respectively. In addition, we 
select galaxies down to $r'=22.2$. This locus in the CMD selects 618 likely galaxy members as 
complete as possible (dashed lines in Fig. \ref{fig:cmr}) in the region considered. 

Fig. \ref{fig:contours}, left panel, shows the contour levels of the isodensity galaxy distribution 
of likely member. This map has been obtained by evaluating the cumulative contribution of 618
Gaussian profiles (with $\sigma=1$ arcsec width) centred on each individual members in a grid of $245 \times 200$ points. 
From this study we can assess that the cluster presents a central high density clump (the main body) 
and the BCG (ID 60) coincident with the highest density peak. A secondary and very dense clump is
toward the south-west and two very BGs (IDs 22 and 21) are positioned very close to its corresponding 
peak. The isodensity contours show a clear elongation from the main body of the cluster to the 
south. This concentration of galaxies shapes a third small substructure toward the south also containing 
a BG (ID 61). And finaly, a fourth galaxy clump, not so clear in the isodensity contours but also 
containing  a BG (ID 93) is located eastward. Thus, we dectect three galaxy clumps (toward the 
south-west, south and east, respectively, and following the order of a decreasing density) 
surrounding a very dense main cluster. Tab. \ref{tab:structures} lists the precise positions and 
global properties for these four significant galaxy clumps.

We remark that the external regions of RXCJ1230 are particularly rich of substructures. The 
substructure configuration reported here using optical data, characterized by a central main body 
with three substructures around it completely agrees with that observed and reported by \citet{Boh22} 
using X-ray data retrieved by XMM-Newton. As Fig. \ref{fig:contours} 
shows, the galaxy density contours (left panel) and X-ray surface brightness profile (right panel) 
are almost coincident and follow the same shape.

\subsection{Spatial-velocity correlations}
\label{sec:spa_vel_correlat}

One of the most useful tools to test the existence of substructures is the study of possible 
spatial-velocity correlations. In this work, we use different techniques to analyse the structure
of RXCJ1230 combining positions and velocities of galaxy members.

The presence of internal structures clearly influences the cluster velocity field. So, in a first 
step, in order to investigate the RXCJ1230 complex, we divide galaxies in two samples. We 
search for possible bimodality through the presence of gaps in the velocity distribution, which 
could cause drops in the galaxy counts for some particular bins of the velocity histogram (see 
as an example Fig. 4 in \citet{Bar07} for a clear bimodal configuration of Abell 773). In our case, 
the most important drop is that detected around -500 km s$^{-1}$ (see Fig. \ref{fig:histogram}, 
inner panel). Considering this drop of galaxy counts around -500 km s$^{-1}$ as a possible frontier 
between two separate galaxy subsamples, we study the spatial distribution of galaxies with v$<-500$ 
km s$^{-1}$ and v$\ge -500$ km s$^{-1}$, respectively. We detect no differences between these two 
galaxy subsets and both populations are homogeneously distributed in space, which suggests that 
this drop in the velocity histogram is not representative of individual galaxy clumps.

\begin{figure}[h!]
\centering
\includegraphics[width=9cm,height=6cm]{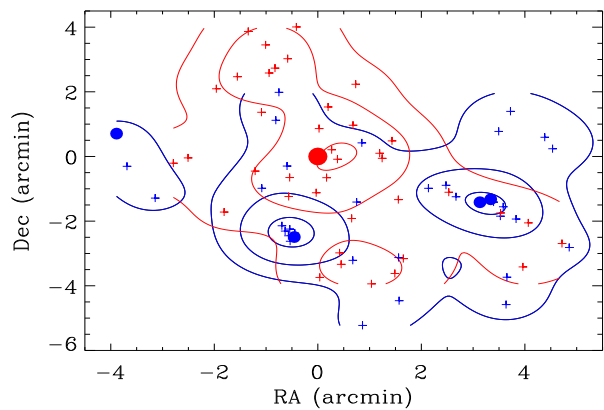} 
\caption{Isodensity contours of spectroscopiacally confirmed galaxy members. Blue contours 
corresponds to the 2D distribution of galaxies with negative velocity with respect to the mean 
one in the cluster rest frame (see Fig. \ref{fig:histogram}, inner panel). Similarly, red contours 
show isodensity levels for galaxies with positive velocity respect the mean cluster velocity. 
This plot is centred on the BCG, marked with a red big dot. Blue dots correspond to 
BGs belonging to the corresponding clumps. Red and blue contours are plotted at the same density level.}
\label{fig:2D_2phist}
\end{figure}

As a second approach, similarly as we proceeded previously, we divide galaxy velocity sample 
in two sets. One subset containing low velocities with $\textrm{v}<\bar{\textrm{v}}$, and a second 
subset with high velocities ($\textrm{v}>\bar{\textrm{v}}$). In other words, the low and high velocity 
subsamples correspond to galaxies with negative and positive velocities respect to the mean one in 
the cluster velocity rest frame (see inner panel of Fig. \ref{fig:histogram}). We check the difference 
between the two distributions of galaxy positions. Fig. \ref{fig:2D_2phist} shows that low and high 
velocity galaxies are segregated, as we advance in Sect. \ref{sec:members_global}. While high velocity 
galaxies (red contours) shape the main body of the system, the low velocity galaxies (blue contours) 
are placed in the three surrounding substructures. This fact can also be noticed in Fig. 
\ref{fig:3plots} (middle panel), where $<$v$>$ takes values about 100200 km s$^{-1}$ for
distances $\sim 0.3-0.4$ Mpc and 99700 km s$^{-1}$ at $>1.2$ Mpc.

We carry out a second test for cheking this spatial-velocity segregation. We combine galaxy
velocities and positions by computing the $\delta$-statistics using the \citeauthor{Dress88} (DS)
test \citep{Dress88}, which looks for groups of galaxies showing deviations from the local velocity mean, 
or with velocity dispersion that differs from the global one. If one assumes a random distribution 
of velocities, one would expect such deviations to be proportional to the number of galaxy members, 
$\Delta \sim N$, for clusters without substructure or clusters with substructures with similar 
velocity dispersion and relative movement in the line of sight with respect their main 
bodies. On the other hand, one would expect $\Delta > N$ for clusters with substructures showing 
relative movements in the line of sight well differenciated from the main body, and/or even subclusters
with lower velocity dispersion respect the main cluster. We find a cumulative deviation of $\Delta=65$ (at 
the 95\% c.l., as estimated by computing 1000 Monte Carlo simulations), which is a value comparable 
to the number of members (77). This suggests that RXCJ1230 presents all its substructures 
moving with no significant deviations in the line of sight component respect the main cluster body 
in agreement with our finding in section \ref{sec:velocity}. In addition, the $\Delta$ value 
shows that no group has been detected with velocity dispersion very different from that of the whole 
cluster. However, the cumulative $\Delta$ is unable to provide information on the presence of possible 
individual galaxy clumps. This issue is explored in the following using the individual $\delta$ associated
to each galaxy position and also obtained through the DS test.

\begin{figure}[h!]
\centering
\includegraphics[width=9cm,height=6cm]{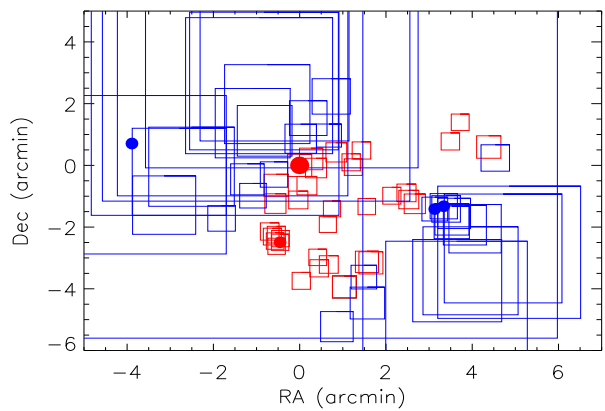}
\caption{Spatial distribution of the 77 cluster members, each marked by a square. The size of 
squares is proportional to $exp(\delta_i)$ computed using the $\delta_i$ deviations obtained in 
the DS test. Red and blue squares separate populations showing deviations lower and higher than 
$\bar{\delta_i}$, respectively. BGC and BGs positions are marked with dots.}
\label{fig:DS_delta}
\end{figure}

Fig. \ref{fig:DS_delta} illustrates the $\delta$-statistics combined with the spatial 
distribution of the 77 cluster members. Around each point a square is plotted with side 
proportional to $exp(\delta_i)$. So, the larger square, the larger is the deviation ($\delta$)
of the local mean velocity from the global mean. Red squares correspond to galaxies
with $\delta_i<\bar\delta_i$ (where $\bar\delta_i$ is the mean $\delta_i$ deviation 
of the 77 cluster members), while blue represents members with deviations $\delta_i>\bar\delta_i$. 
This plot shows that the most important substructures is mainly located toward the east and 
south-west from the main body of the cluster. The galaxies of these groups present higher deviations than 
galaxies in the main body (center). The main advantage of this method is that no $a$ $priori$ selections or assumptions 
about the positions of subclumps have to be imposed. Therefore, this finding is more parameter
independent and so more robust.

To summarize, the study of the individual $\delta$ of the DS test provides us with a 
valuable proof of the presence of substructures in the periphery of the cluster supported by high 
deviations from the mean velocity of the cluster in the external zones. This finding is in  
agreement with what we find in the 2D spatial distribution (see Sect. \ref{sec:spatial_distrib}). 
In addition, the cumulative $\Delta$ is comparable to the number of cluster members, which suggests 
that no large relative movement of these substructures are expected in the radial component. Therefore, 
in agreement with what we find in Sect. \ref{sec:velocity}, relative movements of subclusters respect 
the main body should be (nearly) contained in the plane of the sky.

\begin{figure}[h!]
\centering
\includegraphics[width=9cm,height=6cm]{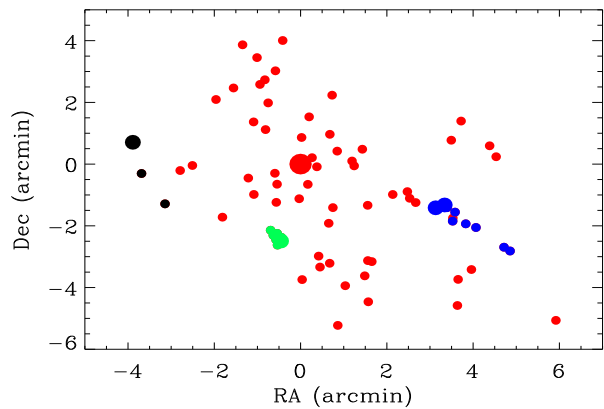}
\caption{Spatial distribution on the sky of the 77 cluster members. The 9 blue and 7
green dots correspond to the galaxies belonging to the south-west and south substructures
identified with $>95$\% probability using the KMM procedure. Red dots
correspond to galaxy members being part of the cluster main body with different 
probabilities following the same algorithm. Black dots correspond to galaxies selected 
manually as part of the eastern clump according to their space-velocity segregation 
(see Fig. \ref{fig:2D_2phist}). Big dots mark the BCG and BGs.}
\label{fig:kmm}
\end{figure}

With the aim of identifying individual galaxies belonging to each substructure, we use a new 3D 
diagnostic test. We apply a 3D version of the Key's Mixture Model algorith \citep[KMM;][]{Ash94} 
in order to separate different components in velocity space. The KMM algorithm estimates the 
probability that a given galaxy belongs to a given component in an iterative procedure. At the 
end, the algorithm produces a list of galaxies associated with the cluster main body and with 
each additional substructure. However, it needs to start from an initial input configuration. 
In order to minimize the dependence of the final results of probabilities on the initial guess 
input, we run KMM with several (more than ten) initial randomly allocations to a three galaxy 
populations, equally composed by 25, 25 and 24 galaxies (the three easternmost galaxies have 
been excluded from this test given the redshift undersampling in this region\footnote{Including 
a fourth galaxy clump with the three easternmost galaxies does not produce a reliable KMM test
result.}). The final result was always the same, converging into a very stable solution. The 
output set of probabilities fits always a three-group partition, assigning 9 and 7 galaxies to 
the south-west and south substructures, respectively, with >95\% probabilities, according 
to the likelihood ratios obtained from KMM test. The rest of the galaxies are 
assigned to the main cluster body. This galaxy allocation, for the main body and the two 
substructures (south-west and south), is shown in Fig. \ref{fig:kmm}. On the other hand, 
the analysis of the $p$-value probability to obtain this KMM result by chance is 
lower than 0.001 ($<0.1$\% probability). The substructure to the east was not 
identified through this technique mainly due to the lack of redshift information in this zone. 
However, the DS test shows high $\delta_i$ deviations in the surroundings of this region, 
around $(-3',1')$ position in Fig. \ref{fig:DS_delta}.

\section{Dynamical mass of RXCJ1230}
\label{sec:dynamics}

On the basis of the previous section, we can conclude that RXCJ1230 is formed by three substructures, 
spatially well separated in the sky. The main substructure is placed to the south-west with 
respect to the main body of the cluster, and contains two very bright galaxies, the BGs-W (1 
and 2). A second substructure is located toward the south, which resembles a very compact group of 
galaxies dominated by a very bright galaxy, the BG-S. In addition, the spatial distribution 
of galaxies and the X-ray surface brightness map suggest also the presence of a third clump 
located to the east and containing a BG. 

\begin{table*}[!htbp]
\caption{Positions and global properties for the whole cluster and the four galaxy clumps detected in RXCJ1230.}
\begin{center}
\begin{threeparttable}[t]
\begin{tabular}{lccccccc}
\midrule \midrule
Structure  & R.A. \& Dec. (J2000)              & N$_{gal}$ & $\bar{\textrm{v}}$ & $\sigma_\textrm{v}$ & $r_{200}$          & $M_{200}$                    & $M_{500}$  \cr
           & R.A.=$12\! : \! mm \! : \! ss.ss$ & & (km s$^{-1}$)  & (km s$^{-1}$)        & ($h_{70}^{-1}$Mpc) & ($\times10^{14}$ M$_{\odot}$) & ($\times10^{14}$ M$_{\odot}$) \cr
           & Dec.=$+34\! :\! ^\prime \! : \! ^\prime$$^\prime$  &  &  &  &  &  &  \cr
\midrule
Global     & 30:45.78  \enspace 39:26.3        & 77 &  99658$\pm$161 & $1004_{-122}^{+147}$ & --         & $14.1 \pm 3.8$  & $9.0 \pm 2.3$ \cr
Centre     & 30:45.78  \enspace 39:26.3        & 58 &  99967$\pm$161 & $999\pm160$          & $\sim1.8$  & $9.0 \pm 1.5$   & $5.6 \pm 1.0$ \cr
South-West & 30:30.55  \enspace 38:01.4        & 9  &  98810$\pm$126 & $792\pm230$          & $\sim1.5$  & $4.4 \pm 3.3$   & $2.7 \pm 2.0$ \cr
East       & 31:04.67  \enspace 40:08.7        & 3  & $\sim 99468$   & $\sim 500$           & $\sim 0.8$ & $\sim 1$        & $\sim 0.7$ \cr
South      & 30:47.98  \enspace 36:56.7        & 7  &  99090$\pm$192 & $<300$               & $<0.5$     & --              & -- \cr
\midrule
% \end{tabular}
% \begin{tablenotes}
% \footnotesize
% \multicolumn{8}{l}{\footnotesize{Note: $\sigma_v$ of the east and south substructures have to be
% taken as guide values, and so the masses and radii derived from them.}}
% \end{tablenotes}
\end{tabular}
\begin{tablenotes}
\item \footnotesize{Note: $\sigma_v$ of the east and south substructures have to be
taken as guide values, as well as the masses and radii derived from them.}
\end{tablenotes}
\end{threeparttable}
\end{center}
\label{tab:structures}
\end{table*}%

Although RXCJ1230 is in a phase of interaction, the main system and the three surrounding
substructures are still well separated and the 2D galaxy density distribution and BGs well
match with their correspondings X-ray peaks. Thus, we can assume that RXCJ1230 is in a 
pre-merging phase. Morover, no shock fronts are detected in the X-ray surface brightness
map, which are very commonly formed after collisions. So, it is reasonable to assume that
each substructure has not yet collided and is roughly in dynamical equilibrium, which allows 
us to compute virial quantities and estimate the mass of the whole cluster as the sum of the 
individual subclump masses.

Table \ref{tab:structures} lists the kinematical properties of this complex, composed of a 
main central cluster surrounded by three substructures. Following the KMM test, we unequivocally
associate 9 and 7 galaxies to the south-west and south substructures (see Fig. \ref{fig:kmm}). 
This allows us to estimate a mean velocity and a rough velocity dispersion to these 
substructures. So, we obtain a $\bar{\textrm{v}}_{SW}=98810 \pm 126$ and $\bar{\textrm{v}}_S=99090 \pm 192$ 
km s$^{-1}$ for the mean velocity of the substructures to the SW and S, respectively. In 
the same way, we estimate a velocity dispersion of $\sigma_{main}=999 \pm 160$, $\sigma_{SW}=792 
\pm 230$ for the main body (assumed to be composed by 58 galaxies) and the south-west clump, 
respectively. The 7 galaxies identified in the southern substructure seem to configure a small 
compact group of galaxies with very low $\sigma_\textrm{v}$. In this case, we could only estimate 
an upper limit of $\sigma_{S}<300$ km s$^{-1}$. Regarding the eastern substructure, we could assume 
that the mean velocity should be similar to its corresponding BG (99100 km $s^{-1}$), and in 
agreement to the low galaxy density and the low X-ray emission observed, the velocity 
dispersion (in this case computed as the $rms$) of these three galaxies is $\sigma_{E} \sim 500$ 
km s$^{-1}$. We obtain velocity dispersion for the east and south substructures as
guide values, and only have to be considered in order to extract information about the
magnitude and the importance of such galaxy clumps in the context of the whole RXCJ1230 cluster.

Given that galaxies are tracers of the gravitational potential of a halo, it is possible to 
estimate the dynamical mass of a system from its velocity dispersion. We use the calculated 
$\sigma_\textrm{v}$ and its relation with $M_{200}$ \footnote{$r_{200}$ is defined as the radius 
inside which the average mass density in the cluster is 200 times the critical density of the 
Universe at the cluster redshift. Similarly to $r_{500}$ for its corresponding mass density. 
Therefore, $M_{200}$ and $M_{500}$ are the virial mass contain within $r_{200}$ and $r_{500}$, 
respectively.} to determine the dynamical mass of RXCJ1230 and its substructures. In the 
literature, there are many scaling relations to obtain dynamical masses of clusters from their 
velocity dispersion. Some examples are that obtained by \citet{Evr08}, \citet{Saro13}, 
\citet{Mun13} and \citet{Ferra20}. All of them produce very similar values for the dynamical 
mass. However, in this work, we follow Munari et al.'s (2013) prescription (see Eq. 1 
therein) given that the relation they obtain is constructed using very complete simulations, 
which take into account not only dark matter particles but also subhaloes, galaxies and AGN 
feedback. Therefore, following \citet{Mun13} $\sigma_\textrm{v}-M_{200}$ relation, we find 
dynamical masses of $M_{200}=9.0 \pm 1.5 \times 10^{14}$ M$_{\odot}$ and $4.4 \pm 3.3 \times 
10^{14}$ M$_{\odot}$ for the main cluster and the south-west substructure, respectively. 
Regarding the eastern substructure, the velocity dispersion estimate is very unaccurate, 
which makes very difficult to determine its mass with a minimum of precision. However, applying 
the same technique we obtain a rough estimate of $\sim 1 \times 10^{14}$ M$_{\odot}$ for the 
eastern clump. In order to compare these values with others in the literature (which mainly 
refer to $M_{500}$), $M_{200}$ can also be converted into $M_{500}$ following the relation 
given by \citet{Duf08}. $M_{500}$ has been rescaled from $M_{200}$ assuming a concentrarion 
parameter $c=3.5$ (an appropriate value for clusters at $z=0.3$ and $M_{200}=10^{14}-10^{15} 
M_{\odot}$), integrating a Navarro-Frenk-White (NFW) profile \citep{NFW97} and interpolating 
to obtain $M_{500}$. So, we obtain $M_{500}=5.6 \pm 1.0 \times 10^{14}$ M$_{\odot}$ and $2.7 
\pm 2.0 \times 10^{14}$ M$_{\odot}$ for the main cluster and the south-west substructure, 
respectively, and a rough (and qualitative) value of $\sim 0.7 \times 10^{14}$ M$_{\odot}$ 
for the eastern clump.

In addition to the mass, we can also estimate the virial radius, $r_{200}$, which provides 
information about the quasi-virialized region, as the radius of a sphere of mass $M_{200}$
and 200 times the critical density of the Universe at the redshift of the system, 200$\rho_c(z)$. 
So, $M_{200}=100 r_{200}^3 H(z)^2/G$. Following this expression we obtain $r_{200}\sim 1.8$
and $\sim 1.5$ $h_{70}^{-1}$ Mpc for the main body and the south-west substructure, respectively. We 
compile the radius and mass estimates in table \ref{tab:structures}.

As for the whole mass of the system, the contribution of the east and south groups is of minor 
importance since they likely have low velocity dispersion and $M_{200}$ scales with 
$\sigma_\textrm{v}^3$. Thus, we can estimate a reliable total mass of the cluster as the sum 
of the main body and the other substructures. So, we obtain a total mass of $M_{200} \simeq 1.4 
\pm 0.4 \times 10^{15}$ M$_{\odot}$.

To summarize, RXCJ1230 is composed of a central main body accreting three substructures
from its enviroment. The main substructure is the one to the south-west, which keeps a mass
ratio about 2:1 respect the main cluster. The substructure to the east is very small, since 
the velocity dispersion and mass estimate show a mass ratio about 10:1 with respect to the main 
body. Something similar may occur with the southern substructure, which seems to be a very 
compact group with very low velocity dispersion. Since we are not able to compute accurate
velocity dispersion for E and S clumps, the values reported in Tab. \ref{tab:structures} for 
these overdensities have to be taken as qualitative results. However, they are important in order
to characterize these minor subclumps. In fact, these findings agree with that obtained by 
\citet{Boh22} from X-ray data, which supports these results.

\subsection{Comparison with X-ray observations}
\label{sec:optical_xray_mass}

By analysing X-ray data, \citet{Boh22} also find a complex configuration of substructures around 
the main body of the cluster, which is elongated toward the south. The main substructure is placed 
to the south-west, while another minor clump is detected toward the east.

The detailed X-ray analysis of the RXCJ1230 complex is presented in \citet{Boh22}. To summarize, 
they find that the X-ray temperature of the main body is $4.7 \pm 0.4$ keV, while the south-west 
and east components show a $T_X=4.4 \pm 0.6$ and $3.3 \pm 0.6$ keV, respectively. On the other 
hand, \citet{Boh22} calculate the mass gas inside $r_{500}$, 3.45, 2.98 and 2.39 arcmin for the 
central, south-west and east cluster components, respectively, and use a $\beta$-model for the 
plasma density distribution. They assume a fix $\beta$-value of 2/3, which is typical for massive 
relaxed clusters. Finally, they adopted several scaling relations with $L_X$, $T_X$, $Y_X$ and 
$M_{gas}$ and of $M_{500}(\beta_{fix}$) and find a total mass of $M_{500,X}=7.7(\pm0.7) \times 
10^{14}$ $M_{\odot}$. 

The sum of dynamical masses corresponding to the main body and south-west clumps, the two main 
systems of RXCJ1230, is $M_{500,dyn}\simeq 9.0 \pm 2.3 \times 10^{14} \ M_{\odot}$. Therefore, comparing 
this value with that derived from X-ray, we see that $M_{500,dyn}$ and $M_{500,X}$ are in agreement
within errors and they differ by about 15\%. 

Comparing the mass of individual clumps, \citet{Boh22} find a $M_{500,X}=3.7 \pm 0.5 \times 10^{14}$ 
$M_\odot$ within $r_{500}$ in the main cluster, while we obtain $M_{500,dyn}=5.6 \pm 1.0 \times 10^{14} 
\ M_{\odot}$. This discrepancy of about 30\% may come from the fact that $M_{500,dyn}$ has been derived 
by converting $M_{200}$ into $M_{500}$. In fact, when considering only the 47 galaxy members with 
redshift within $r_{500}$ (=3.45 arcmin =1 Mpc; see Tab. 3 in \citeauthor{Boh22} \citeyear{Boh22}), a 
$\sigma_\textrm{v}=913 \pm 133$ km s$^{-1}$ is obtained, which points to a mass of about $M_{500,dyn} 
\sim 4.3\times 10^{14} \ M_{\odot}$, almost coincident with that derived from X-ray data. For the 
south-west component we find a good agreement of the mass determination with $M_{500,X} = 2.5 \pm 0.64 
\times 10^{14}$ M$_{\odot}$ from X-rays and $M_{500,dyn} = 2.7 \pm 2 \times 10^{14}$ M$_{\odot}$ from 
the galaxy dynamics. For the eastern component a mass estimate is more difficult. However, even 
considering rough estimations, $M_{500,dyn} \sim 1 \times 10^{14}$ M$_{\odot}$ and $M_{500,X} = 1.35 
\pm 0.3 \times 10^{14}$ M$_{\odot}$ roughly agree within errors.

\section{Dynamics and merging}
\label{sec:merging}

As we pointed out above, both the main cluster and the three substructures are well detectable 
and optical and X-ray data indicate very similar locations, so we are looking at RXCJ1230
in a prior merging phase. However, the velocity distribution and the not so well resolved 
galaxy populations suggest that the substructures start to interact with the main cluster.
With this scenario, the main collision will be produced between the substructure to the
south-west involving a mass ratio of 2:1. In fact, the X-ray temperatures of the main body and
the substructure are very similar, $T_{X,C}=4.7 \pm 0.4$ and $T_{X,SW}=4.4 \pm 0.6$ keV 
\citep{Boh22} for the central main clump and the south-west substructure, respectively, even 
if they show very different dynamical masses. % and X-ray luminosities ($L_{X,C}=2.12 \pm 0.05$ 
% and $L_{X,SW}=0.71 \pm 0.14$ in the 0.5-2 keV energy band). 
Whereas typical X-ray temperatures of relaxed clusters with $M_{500} \sim 2.7 \times 
10^{14}$ M$_{\odot}$ are about 3 keV \citep[see e.g. Fig.9 in][and references therein]{ket13}, 
the south-west substructure, could be starting to show an enhancement of its intra-cluster 
medium (ICM) temperature. This enhancement supports the fact that the main body and 
south-west substructure are starting to collide.

When the merging scenario is assumed to explain an enhancement of the ICM temperature, a 
relative colliding velocity is needed to heat up the ICM \citep{Gut05}. On the assumption 
that the two components are to cause a head-on collision and that their kinetic energies are 
completely converted to thermal energy, the colliding velocity is $v_{coll}^2=3 k \Delta T / \mu 
m_p$ km s$^{-1}$, following prescriptions detailed in \citet{Shi99}, where $\mu$ and $m_p$ are 
the mean molecular weight (0.6) in amu, and the proton mass, respectively. So, assuming an excess 
temperature of $k \Delta T \sim 1.5$ keV, we find a $v_{coll} \simeq 850$ km s$^{-1}$, 
which is in very good agreement with the observed relative LOS velocity in the cluster rest 
frame, as computed from $(\bar{\textrm{v}}_C - \bar{\textrm{v}}_{SW} )/(1+z)= 869$ km s$^{-1}$.

\subsection{Two body merging model}
\label{sec:bimodel}

In this section, we investigate the relative dynamics of RXCJ1230 main body (C) and its main
substructure, the south-west (SW) one, which dominates the dynamics of the whole cluster, with 
a mass ratio about 2:1. The rest of the interactions, including the south and east substructures, 
are minor merger events with mass ratios about 10:1. We analyze this C-SW interaction from 
different approaches, which are based on an energy integral formalism in the framework of
locally flat spacetime and Newtonian gravity \citep[see e.g.][]{Bee82}.
The three relevant observable quantities for the two system interaction are: the relative 
line-of-sight velocity, $V_r=869 \pm 153$ km s$^{-1}$; the projected physical distance, $D=0.97$ 
$h_{70}^{-1}$ Mpc (3.38 arcmin); and the total mass of the two systems by adding the masses of 
the two clumps within $r_{200}$, $M_{sys} \sim 13.4 \pm 3.6 \times 10^{14}$ M$_{\odot}$ (see Sect. 
\ref{sec:dynamics}).

\begin{figure}[h!]
\centering
\includegraphics[width=9cm,height=6cm]{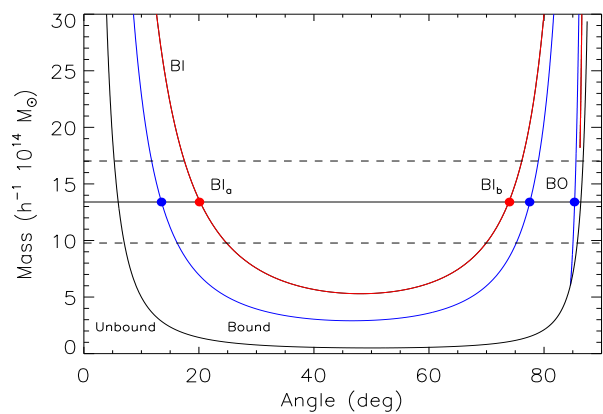}
\caption{Two-body model applied to the main cluster and south-west galaxy substructure
(C-SW system). The black curve separates bound and unbound regions according to the Newtonian 
criterion. Solid curves represent the bound incoming (BI) and bound outgoing (BO) solutions. 
Blue and red curves denote the models for 706 and 1022 km s$^{-1}$, which represent the 
marginal relative velocity between main cluster and substructure, considering the 
corresponding uncertainties. The horizontal lines represents the observational values
of the total mass of the C-SW system, with it uncertainty (dashed lines).}
\label{fig:bimodal}
\end{figure}

First, we consider the Newtonian criterion for the gravitational binding, which follows
the expression $V_r^2 D \leq 2GM_{sys} \sin^2 \alpha \cos \alpha$, where $\alpha$ is the 
projection angle between the line connecting the centres of the two clumps and the 
plane of the sky. Fig. \ref{fig:bimodal} represents the two-body model obtained. Considering
the value of $M_{sys}$, the C-SW system is bound between 6$^\circ$ and 86$^\circ$ with
a probability of $\int_{6^\circ}^{86^\circ} \cos \alpha d \alpha = 0.89$ (i.e. 89\%).

Then, we apply the analytical two-body model introduced by \citet{Bee82} and \citet{Tho82} 
\citep[see also][]{Lub98}. This model assumes radial orbits and not rotation of the system. 
In addition, the clumps are assumed to start their evolution at $t_0=0$ with a separation of 
$d_0=0$ and are moving apart or coming together for the first time in the history. That is, 
with this model, we are assuming that we are seeing the cluster prior to collision (at $t=9.98$ 
Gyr at the redshift of RXCJ1230). The solutions for this model is shown in Fig. \ref{fig:bimodal}, 
where we compare the total mass of the system, $M_{sys}$, with the projection angle, $\alpha$. 
The possible solutions span several cases: two bound incoming solutions (BI$_a$ and BI$_b$) 
around 14$^\circ$-20$^\circ$ and 74$^\circ$-77$^\circ$, respectively, and one bound outgoing (BO) 
at $\sim 85^\circ$. The incoming case is degenerated because of the ambiguity in the projection 
angle $\alpha$. So, for simplicity, we assume the mean value in the incoming cases, which is
BI$_a\sim 17^\circ$ and BI$_b\sim 76^\circ$ and that the region of $M_{sys}$ values are equally 
probable for individual solutions. Under these assumptions we estimate the following 
probabilities: $P_{BIa}\sim 87$\%, $P_{BIb}\sim13$\% and $P_{BO}\ll 0.1$\%. We discard the BO 
solution because it is very unlikely and we analyse below the BI ones.

Between the two possible incoming solutions, $\alpha=17^\circ$ and $\alpha=76^\circ$, the second
one is quite unlikely. An $\alpha=76^\circ$ (associated to the BI$_b$ solution) would imply 
a distance between clumps of $\sim 4.0$ $h_{70}^{-1}$ Mpc, which more than twice $r_{200}$. On 
the other hand, an $\alpha=17^\circ$ (BI$_a$) is the most likely solution and, for this case,
the distance C-SW centres would be $\sim 1.0$ $h_{70}^{-1}$ Mpc, which would explain a certain
degree of interaction. Thus, when assuming $\alpha=17^\circ$ the colliding velocity would 
be $\sim 3000$ km s$^{-1}$ and the cluster clumps will cross after $\sim 0.3$ Gyr.

For sure, the characterization of the dynamics of RXCJ1230 through these models is affected by 
several limitations. First, the two-body model does not consider the possibility of an off-axis
merger neither a mass distribution in the subclusters. And secondly, this study does not 
take into account the presence of minor subclusters (the south and east clumps). Therefore, the
model here presented has to be taken as one of the many possible approaches to the collision
scenario in RXCJ1230.

\section{Summary and conclusions}
\label{sec:conclusions}

We present the results of the kinematical and dynamical state of the complex galaxy cluster 
RXCJ1230.7+3439. Our study is based on new spectroscopic redshifts acquiered at the 3.5m TNG 
telescope covering a region of $\sim 8^\prime\times 8^\prime$. We also consider some SDSS DR16 
spectroscopic redshifts in order to complement our sample. In addition, we use the SDSS 
photometry in a field of $\sim 13^\prime\times10^\prime$ to analyse the spatial distribution of 
likely cluster members.

We select 77 galaxy cluster members around $z=0.332$ and compute a LOS global velocity 
dispersion of $\sigma_\textrm{v}=1004_{-122}^{+147}$ km s$^{-1}$.

Our analysis confirms the presence of three substructures surrounding the main body of the 
cluster and they are well recognizable in the plane of the sky in the 2D spatial distribution 
of galaxies. Moreover, we identify several bright galaxies dominating the core of each 
substructure. The more massive substructure is the one located to the south-west, which shows 
a $\sigma_\textrm{v} \sim 800$ km s$^{-1}$ and differs by $\sim 870$ km s$^{-1}$ from the main 
body in the LOS velocity. The southern substructure resembles a compact group of galaxies with 
very low velocity dispersion ($\sigma_S < 300$ km s$^{-1}$), while we find that the eastern 
substructure shows a $\sigma_E \sim 500$ km s$^{-1}$. The dynamical masses estimated from 
these velocity dispersions are $M_{200}=9.0 \pm 1.5 \times 10^{14}$ M$_{\odot}$, $4.4 \pm 3.3 
\times 10^{14}$ M$_{\odot}$ and $\sim 1 \times 10^{14}$ M$_{\odot}$ for the main cluster, the 
south-west and east substructures, respectively. Thus, considering the complex structure of 
RXCJ1230, we estimate that the whole cluster contains a total mass in the range of $M_{200} 
\simeq 1.4 \pm 0.4 \times 10^{15}$.

Given that the galaxy density peaks coincide with those observed in the X-ray surface
brightness, we infer that the system agrees with a pre-merging event, where the main
collision is that involving the main body and the south-west clump. This interaction occurs 
with a mass ratio of 2:1 and an impact velocity of $\Delta v_{rf}\sim 3000$ km s$^{-1}$. The 
most likely solution obtained from a two-body problem for these two systems suggests that the 
merging axis lies at $\sim 17^\circ (\pm3^\circ)$ with respect to the plane of the sky and the 
systems will be completely joined in about 0.3 Gyr. However, a slightly increase of the X-ray 
temperature ($k \Delta T \sim 1.5$ keV) in the south-west may indicate that we are observing a 
certain degree of interaction.

\begin{acknowledgements}
\\
We thank to the referee for his useful comments, which have really helped the authors 
to improve this work. \\

R. Barrena acknowledges support by the Severo Ochoa 2020 research programme of 
the Instituto de Astrof\'{\i}sica de Canarias. H. B\"ohringer acknowledges support 
from the Deutsche Forschungsgemeinschaft through the Excellence cluster "Origins". 
G. Chon acknowledges support by the DLR under the grant n$^\circ$ 50 OR 1905. \\

This article is based on observations made with the Italian Telescopio Nazionale 
Galileo operated by the Fundaci\'on Galileo Galilei of the INAF (Istituto Nazionale 
di Astrofisica). This facility is located at the Spanish del Roque de los 
Muchachos Observatory of the Instituto de Astrof\'{\i}sica de Canarias on the 
island of La Palma. \\

Funding for the Sloan Digital Sky Survey (SDSS) has been provided by the Alfred
P. Sloan Foundation, the Participating Institutions, the National Aeronautics
and Space Administration, the National Science Foundation, the U.S. Department
of Energy, the Japanese Monbukagakusho, and the Max Planck Society. \\

The Pan-STARRS1 Surveys (PS1) and the PS1 public science archive have been made 
possible through contributions by the Institute for Astronomy, the University of 
Hawaii, the Pan-STARRS Project Office, the Max-Planck Society and its participating 
institutes, the Max Planck Institute for Astronomy, Heidelberg and the Max Planck 
Institute for Extraterrestrial Physics, Garching, The Johns Hopkins University, 
Durham University, the University of Edinburgh, the Queen's University Belfast, 
the Harvard-Smithsonian Center for Astrophysics, the Las Cumbres Observatory 
Global Telescope Network Incorporated, the National Central University of Taiwan, 
the Space Telescope Science Institute, the National Aeronautics and Space 
Administration under Grant No. NNX08AR22G issued through the Planetary Science 
Division of the NASA Science Mission Directorate, the National Science Foundation 
Grant No. AST-1238877, the University of Maryland, Eotvos Lorand University (ELTE), 
the Los Alamos National Laboratory, and the Gordon and Betty Moore Foundation.
\end{acknowledgements}

\bibliographystyle{aa}

\end{document}